\documentclass[11pt]{article}
\usepackage{graphicx}
\usepackage{setspace}
\usepackage{color}
\usepackage[super,sort&compress]{natbib}
\usepackage{multirow}
\usepackage{bm}
\usepackage{amsfonts}
\usepackage{amssymb}
\usepackage{enumerate,ulem}
\usepackage{float}
\usepackage{ulem} 
\usepackage{color}

\newcommand{\DyT}{Dy$_{2}$Ti$_{2}$O$_{7}$}
\newcommand{\HoT}{Ho$_{2}$Ti$_{2}$O$_{7}$}
\begin{document}

\begin{center}

{\bf Experimental signature of the attractive Coulomb force between positive and negative magnetic monopoles in spin ice}

{C. Paulsen,$^{1 \ast}$ S. R. Giblin,$^{2}$ E. Lhotel,$^{1}$  D. Prabhakaran,$^{3}$\\
G. Balakrishnan$^{4}$,
 K. Matsuhira,$^{5}$ S. T. Bramwell.$^{6}$\\
\normalsize{$^{1}$Institut N\'{e}el, C.N.R.S - Universit\'e Joseph Fourier, BP 166, 38042 Grenoble, France.}\\
\normalsize{$^{2}$School of Physics and Astronomy, Cardiff University, Cardiff, CF24 3AA, United Kingdom.}\\
\normalsize{$^{3}$Clarendon Laboratory, Physics Department, Oxford University,}\\ 
\normalsize{Oxford, OX1~3PU, United Kingdom.}\\
\normalsize{$^{4}$Department of Physics, University of Warwick, Coventry, CV4 7AL, United Kingdom}\\
\normalsize{$^{5}$Kyushu Institute of Technology, Kitakyushu 804-8550, Japan.}\\
\normalsize{$^{6}$London Centre for Nanotechnology and Department of Physics and Astronomy,}\\
\normalsize{ University College London, 17-19 Gordon Street, London, WC1H 0AJ, United Kingdom.}\\
\normalsize{$^\ast$ E-mail:  carley.paulsen@grenoble.cnrs.fr}
}

\end{center}

\vspace{0.5cm}

{\bf  A non-Ohmic current that grows exponentially with the square root of applied electric field is well known from thermionic field emission (the Schottky effect)~\cite{Schottky}, electrolytes (the second Wien effect) \cite{Onsager} and semiconductors (the Poole-Frenkel effect)~\cite{Frenkel}. It is a universal signature of the attractive Coulomb force between positive and negative electrical charges, which is revealed as the charges  are driven in opposite directions by the force of an applied electric field.  Here we apply thermal quenches~\cite{Paulsen} to spin ice~\cite{Harris,BramwellHarris,Ramirez,Siddharthan,denHertog,Melko,BramwellGingras} to prepare metastable populations of bound pairs of positive and negative emergent magnetic monopoles~\cite{CMS,Ryzhkin,Jaubert1,Jaubert2,quench} at millikelvin temperatures. We find that the application of a magnetic field results in a universal exponential-root field growth of magnetic current, thus confirming the microscopic Coulomb force between the magnetic monopole quasiparticles and establishing a magnetic analogue of the Poole-Frenkel effect. At temperatures above 300 mK, gradual restoration of kinetic monopole equilibria causes the non-Ohmic current to smoothly evolve into the high field Wien effect~\cite{Onsager} for magnetic monopoles, as confirmed by comparison to a recent and rigorous theory of the Wien effect in spin ice~\cite{Nmat, Kaiser}. Our results extend the universality of the exponential-root field form into magnetism and illustrate the power of emergent particle kinetics to describe far-from equilibrium response in complex systems.}

\newpage
 
Spin ices like \HoT~and \DyT~are almost ideal ice-type or 16-vertex model magnets, embellished by dipole-dipole interactions~\cite{Harris,BramwellHarris,Ramirez,Siddharthan,denHertog,Melko,BramwellGingras}. These long-range interactions are self-screened in the ground state~\cite{Melko} but survive in excited states, where they transform to a Coulomb interaction between emergent magnetic monopole quasiparticles~\cite{CMS,Ryzhkin}. Spin ice may be represented as a generalised Coulomb gas in the grand canonical ensemble, with for \DyT, a chemical potential of $\mu \approx -4.35$ K, set by the original Hamiltonian parameters~\cite{CMS,Ryzhkin,Jaubert1,Jaubert2}. 
Monopoles are thermally generated in dipole pairs which subsequently fractionalise to form free monopoles. At equilibrium, free monopoles coexist with a thermal population of bound pairs that are closely
analogous to Bjerrum pairs in a weak electrolyte~\cite{BramwellGiblin,Giblin}, or more generally analogous to excitons in a semiconductor. Monopole generation and annihilation is thus represented by the following scheme of coupled equilibria:
\begin{equation}\label{eq}
(0) = (+-) = (+) + (-),
\end{equation}
where $(0)$ denotes the monopole vacuum,  $(+-)$ denotes the bound pairs and $(+)$ and $(-)$ denote the free charges. The reactions that make up this scheme define an emergent particle kinetics that may be used to calculate dynamical quantities that depend on the monopole density. 

At temperatures below 0.6 K, the spin degrees of freedom of  \DyT~ gradually fall out of equilibrium~\cite{Schiffer} and spin ice enters a state with the residual Pauling ice entropy~\cite{Ramirez}. The entropy may diminish on exceptionally long time scales ($\geq 10^6$ s) suggesting an approach to an ordered~\cite{Pom} or quantum spin liquid state~\cite{Shannon_new}, but this physics is irrelevant here. We address ordinary experimental time scales, where spin ice is of great interest as a model non-equilibrium system.  

From Maxwell's equations, the current density of magnetic monopoles (`magnetricity') is the rate of change of sample magnetisation, ${\bf J}=\partial {\bf M}/\partial t$. The generalised thermodynamic force that drives the monopole current~\cite{Ryzhkin,PhilTrans} is ${\bf H} - \mathcal{D} {\bf M} - {\bf M}/\chi$, where ${\bf H}$ is applied magnetic field, $- {\bf M}/\chi$ is an entropic reaction (`Jaccard') field and $\mathcal{D} {\bf M}$ is the demagnetising field. 
To simplify the analysis and avoid problematic demagnetising corrections~\cite{Bovo_demag} we work at very small magnetisation. This ensures that the Jaccard and demagnetising fields are negligible and that the monopole conductivity becomes $\kappa = J/H$, analogous to the conductivity of an electrolyte  (= current density/electric field). 
Our experiments are conducted in the low temperature regime of very dilute monopoles, where it is expected~\cite{Ryzhkin} that $\kappa$ is a measure of the instantaneous monopole density, $\kappa \propto n$. Further details, and a discussion of the definition of conductivity, are given in the Supplementary Information (SI1 and SI2 respectively).  

The loss of magnetic equilibrium at $ \sim0.6$ K is connected to the rarifaction of the monopole gas. The equilibrium monopole density decreases with temperature as $n_0 \sim \exp(\mu/T)$ (where $\mu \approx -4.35$ K as above).  Spin flips correspond to monopole hops, so a finite monopole density is required to mediate the spin dynamics. Therefore, at our base temperature, 65 mK, close-to equilibrium relaxation will be exceedingly small and very difficult to measure. A stratagem to avoid this fate, and to ensure access to non-equilibrium behaviour, is to use fast thermal quenches to prepare the sample with a significant density of frozen in, `non-contractable' monopole-antimonopole pairs~\cite{quench} as well as, perhaps, a small density of free monopoles.  In a previous work \cite{Paulsen}  we demonstrated that the magneto-thermal  Avalanche Quench protocol (AQp) results in the fastest thermal quench, giving a very large and \textit{ reproducible} non-equilibrium density of defects, or monopoles, at very low temperature. In Ref.\citenum{Paulsen} the time-dependent magnetisation at relatively long times was qualitatively interpreted by a non-interacting monopole theory. Here we investigate the short-time limit, where a strong effect of monopole interactions is expected (see below).  We exploit the AQp as well as the Conventional Cooling protocol (CCp) of Ref. \citenum{Paulsen}, to quantitatively determine the initial monopole current and monopole interactions.

To this end, we made extensive low temperature magnetisation measurements on 
three different single crystals of \DyT{} prepared at different facilities (labelled 1-3, see Fig.1 and SI1). One of the samples (1) was measured along two different field orientations, while another (3) had the $\sim 10$\% nuclear spins removed. The results were essentially the same in each case, so for simplicity, we describe results for a single sample in a particular orientation: sample (1), a flat ellipsoidal crystal, with the field applied along the long [111] axis.
Crucial to the present investigation was our ability to change the magnetic field at a rate of 1.8 T.s$^{-1}$ (where T = tesla), and to make reliable measurements at the instant the field attains the target value. For these extremely detailed and nonstandard low temperature measurements (see SI1), we used a SQUID magnetometer that was developed at the Institut N\'{e}el in Grenoble. It was equipped with a miniature dilution refrigerator capable of cooling the sample to 65 mK. 

Fig. 1a and b shows how the magnetisation at 65~mK evolves with time and applied field following AQp and a 360 second wait period before the field application (details in SI1). During the short time that the field is being ramped up to the target value, the data points are spurious (shown as the grey area in Fig. 1b), but for longer times the data points are dependable. After the field change the magnetisation $M$ grows precipitously, suggesting that frozen monopole pairs dissociate and separate, forming chains of overturned spins.  The magnification of rapid growth period (Fig. 1b) shows polynomial fits to the data (disregarding the spurious points) from which the magnetic current  density $J = \partial M/\partial t$ was evaluated at  the moment the field has reached its target value.  Two examples are shown as straight lines in Fig. 1b and $J(H)$ at 65~mK is plotted as a function of magnetic field in Fig. 1c. Note that an initial jump in $M$, clearly seen in the figure, occurs while the field is being energized. This jump is very small: $< 0.3$ $\%$ of the final equilibrium magnetisation. Most of the jump can be accounted for by a temperature and frequency independent adiabatic susceptibility. It is discussed in detail in SI1 and is not considered further here. 

The effect of the quench rate and wait time is shown in Fig. 1d, where we plot  $\log J$  vs $\sqrt{H}$ at $T=65$~mK for three different sample preparations (i.e. AQp or CCp with a 2 h or 360 s wait). As anticipated, $J$  depends on the quenched monopole density with AQp resulting in a larger monopole current than CCp. However, the effect of wait time shows very clearly that even at very low temperatures, quenched monopoles are not dynamically frozen, but apparently can still hop and recombine, reducing the density, and resulting in lower monopole current for longer wait times. Yet, all the curves shown in Fig. 1d are qualitatively  similar:  the current density shows a linear increase at small fields, corresponding to Ohmic conduction, followed by a dramatic non-Ohmic growth at larger fields.  In the inset of Fig. 1c we plot the high field ($\mu_0 H\ge0.02$ T) data points to determine the exponent $\alpha$ of $H$ in $J \sim \exp(H^\alpha)$ and find $\alpha \approx 1/2$.

As emphasised above, the $\exp(\sqrt{H})$ behaviour is a defining characteristic of conduction in Coulomb gases where the deviation from Ohm's law arises from the field-induced unbinding of microscopic charge pairs~\cite{Schottky,Onsager, Frenkel}. In spin ice (Fig. 1c)  it is natural to associate the initial Ohmic current with quenched free monopoles and the non-Ohmic current with field-induced unbinding of non-contractable pairs. 
 The analogy with the Poole-Frenkel effect~\cite{Frenkel} is particularly apt as the latter is often associated with field assisted ionisation of metastable traps. 
For a simple heuristic derivation of the exponential-root field limiting form, 
consider a positive and negative monopole at coordinates $\pm r/2$ respectively, that are bound by their mutual Coulomb attraction and dissociate under the influence of a magnetic field. The potential energy of the pair is: 
\begin{equation}\label{energy}
E(r) =  \frac{-\mu_0 Q^2 }{4 \pi}  \frac{1}{r} -\mu_0 Q H r
\end{equation}
where $Q$ is the monopole charge and $\mu_0$ the vacuum permeability. The maximum in the potential energy is at  $r_0 = \sqrt{Q / 4 \pi H}$ and the field lowers the Coulomb barrier to dissociation by an amount
$\Delta E = \sqrt{\mu_0^2 Q^3 H / \pi}$. The rate of escape over this barrier and hence the current becomes:
\begin{equation}\label{escape_rate}
J \propto  \exp (\Delta E/kT) = \exp \sqrt{\beta_C H}, 
\end{equation}
where $\beta_C = \mu_0^2 Q^3/\pi k^2 T^2$ and $k$ is Boltzmann's constant.  As shown in SI3, the amplitude $\beta_C$ more generally depends on the detailed calculation and physical characteristics of the escape process, but the $\exp({\sqrt{H})}$ form is robust to such details. It is a distinctive characteristic of the Coulomb force law, rather than any other pair interaction. 
 
Our observation of the non-Ohmic current at base temperature thus supports two of the most basic predictions of the monopole model: that defects in the spin ice state interact by Coulomb's law~\cite{CMS} and may be trapped in metastable pairs following a thermal quench~\cite{quench}. We proceed to test a third basic expectation, that as the temperature is raised from 65 mK there should be a gradual restoration of the kinetic equilibria between the monopole vacuum, bound pairs and free monopoles (Eqn. \ref{eq}). This should lead~\cite{BramwellGiblin,Giblin,Kaiser} to the appearance of the second Wien effect, the remarkable field-assisted density increase, first understood by Onsager~\cite{Onsager}. 

Conductivity versus field curves are shown in Fig. 2a. We immediately note three qualitative features that are strongly characteristic of the Wien effect. The first is an increase of conductivity with temperature as the monopole states are thermally populated. The second is a crossover from Ohmic conductivity at low field, caused by charge screening, to non-Ohmic conductivity at high field, caused by the field sweeping away the Debye screening cloud~\cite{Onsager}. The third is a crossing of the curves as a function of field and temperature, which may arise from the competing effects of the zero field charge density being an increasing function of temperature and the Wien effect being a decreasing function of temperature. Given these qualitative features it is appropriate to fit our data to theoretical expressions for the Wien effect for magnetic monopoles~\cite{Kaiser}. In SI4 we give a detailed analysis of the applicability of the theory of Ref.~\citenum{Kaiser} to our experiment. 
 
As a first approach (Fig. 2b), we fit the high field conductivity to the Onsager expression 
$\kappa = \kappa_0 \sqrt{F(H)}$ 
where $F(H)$ is unity in zero field and varies as $\exp(\sqrt{H})$ in high field -- see Methods and SI3. Here the parameter $\kappa_0$ represents the zero field conductivity of the ideal  (unscreened) lattice gas~\cite{Onsager,Nmat, Kaiser}. It was treated as an adjustable parameter along with the charge $Q$ (which enters into $F$)  -- hence two parameters were estimated from fits to the data. Excellent fits were obtained (Fig. 2b - the field range used for the fits is discussed in Methods and SI1).
The fitted parameters $\kappa_0(T)$ and $Q_{\rm exp}(T)$ are shown in Fig. 2c. As expected, at lower temperatures, the Wien effect fits return values of the charge that are quite far from the theoretical value. However, at $T \geq 0.3$ K,  the estimated monopole charge is very close to the theoretical one and $\kappa_0(T)$ is consistent with the theoretical expression $\kappa_0 = \nu_0 \exp(-4.35/T)$, with $\nu_0 \approx 300$ s$^{-1}$ (Fig. 2c). Hence our data is consistent with the expected restoration of monopole kinetic equilibria as the temperature is raised above 300 mK. 

An alternative is to treat the temperature $T$ {in the Onsager function (Methods)  as an adjustable parameter, in place of the charge $Q$. The result, Fig. 2d, shows how the fitted parameter $T_{\rm eff}(T)$ tracks the set temperature at $T > 0.3$ K, but becomes roughly constant at lower temperatures. A finite wait time or slower cooling, results in the experimental temperature becoming closer to the set value, suggesting a return to equilibrium (Fig. 2d). Further work is needed to decide if such an effective temperature has any physical meaning in characterising this non-equilibrium system. 

A much more stringent test of the theory~\cite{Kaiser} is to fit the whole conductivity versus field curve, rather than just the high field part. Theoretically, in the dilute limit, the ratio $\kappa/\sqrt{F}$ is simply $\kappa_0(T)$, the conductivity of the ideal lattice gas. However at finite density, charge screening shifts the equilibrium such that, in zero field, $n$ and hence $\kappa$ are elevated by a factor of $1/\gamma_0$, where $\gamma_0<1$ is the Debye-H\"uckel activity coefficient. A strong applied field `blows away' the screening cloud and the correction disappears. An exponential decrease of $\kappa/\sqrt{F}$ from its zero field value $\kappa_0/\gamma_0$ to its limiting high-field value $\kappa_0$, has been confirmed numerically~\cite{Nmat}. The decay rate is determined only by $\gamma_0$ (see Methods). 

To test this, we plot $\kappa({\rm experimental})/\sqrt{F}$ (see Methods) against applied field in Fig. 3a, where the ratio is seen to behave qualitatively as expected, approaching a constant $\kappa_0(T)$, at high field (confirming the high field Wien effect), and rising exponentially to a higher value in zero field. Determining $\kappa_0$ and $\gamma_0$ from the limiting values allows us to compare the predicted exponential crossover with experiment. There is relatively close agreement, (including reproduction of the non-monotonic $\kappa(H)$, Fig. 3b), with $\kappa_0$ showing its theoretical temperature dependence $\kappa_0 \propto \exp(-4.35/T)$ (above 0.3 K), as in Fig. 2c. However, the estimated $\gamma_0 \approx 0.1$ is much smaller than the theoretical value from Debye-H\"uckel theory,  $\sim 0.75$: the origin of this quantitative discrepancy needs further investigation. 
The corresponding current, $J(H)$, is shown in  Fig. 3c, where the deviations from Ohm's law, and the general success of the theory, are clearly illustrated. Fig. 3d illustrates the effect of waiting for 2 h before the measurement, demonstrating that the functional form of $J(H)$ is largely independent of the initial monopole concentration. 
  
In SI5 we show how our experimental measurements in low field are consistent with previous magnetic relaxation~\cite{Giblin} and alternating current (ac)susceptibility measurements~\cite{Mats,Yaras}. The relevance of material defects to magnetic relaxation in spin ice has been explored~\cite{Revell,Goff}, but such subtle near-to-equilibrium effects are outside the resolution of our experiment which explores the intrinsic far from equilibrium response. Finally, the first report of the low-field Wien effect for magnetic monopoles in spin ice, based on a muon method ~\cite{BramwellGiblin}, led to controversy~\cite{Dunsiger, Blundell,Lees} which is as yet unresolved -- see Ref.~\citenum{Nuccio} for a review. In contrast, our more direct method unambiguously establishes the spectacular high-field Wien effect for magnetic monopoles.

The monopole model provides an essentially complete analysis of the far-from equilibrium magnetisation in spin ice and shows how a model non-equilibrium system may be understood in terms of quasiparticle kinetics (see Fig. \ref{four}).  The monopole conductivity (which in magnetic language is a time-dependent susceptibility) displays a field-dependence different to that of a paramagnet (typically $constant + O(H^2)$~\cite{SK}) or a spin glass (typically a weakly decreasing function of field~\cite{Tholence,Lundgren}). Our result illustrates the advantages of mapping a complex far from-equilibrium system on to a weak electrolyte. 
The emergent particle kinetics means that testable predictions may be made about spin relaxation, even when the initial state of the system is not known. It would be interesting to see if this approach can be generalised beyond spin ice to other complex magnets and other glassy systems. 

\vspace{0.5 cm}
\noindent
{\bf Methods}

Data for $\mu_0 H \ge 0.15 $ T are excluded as they suffer excessive quasi-Joule heating (an issue in analogous electrolytes), while points at $\ge$ 0.3 K, $> 0.12$ T are displayed but not fitted, to guard against systematic error: see SI1 for details. Onsager's function is
$F(b) =I_1\left(\sqrt{8b}\right)/\sqrt{2b}$, where $I_1$ is the modified Bessel function and $b = \mu_0^2 Q^3 H/8 \pi k^2 T^2$. In Fig. 3 $\tilde{\kappa}(H) \equiv \kappa(H)/\sqrt{F(H)}$ is fitted to~\cite{Kaiser}
$
\tilde{\kappa}(H) = \kappa(0) \left[ \gamma_0 + (1-\gamma_0) \exp \left(-\frac{\mu_0^2Q^3 H(1-\gamma_0)}{16 \pi k^2 T^2}  \right) \right]
$
- see SI4 - where $Q = 4.20132 \times10^{-13}$ Am (theoretical charge calculated from a more accurate moment and lattice parameter than used in Ref.~\citenum{CMS}).

{\bf Data Accessibility.} The underlying research materials can be accessed at the following:\newline http://dx.doi.org/10.17035/d.2016.0008219696. 

\vspace{0.5cm}
\noindent
{\bf Acknowledgements} 

CP acknowledges discussions and mathematical modelling help from Claude Gignoux. STB thanks his collaborators on Refs. \citenum{Nmat,Kaiser} -- Vojt\v{e}ch Kaiser, Roderich Moessner and Peter Holdsworth  --  for many useful discussions concerning the theory of the Wien effect in spin ice. SRG thanks EPSRC for funding. We thank Martin Ruminy for assistance with sample preparation.  

\vspace{0.5cm}
\noindent
{\bf Author Contributions} 

Experiments were conceived, designed and performed by C.P., E.L. and S.R.G.. The data were analyzed by C.P., E.L., S.R.G. and S.T.B., who adapted the theory of Ref. \citenum{Kaiser}.  Contributed materials and analysis tools were made by K.M., D.P. and G.B.. The paper was written by S.T.B., C.P., E.L. and S.R.G..

\newpage

\begin{figure}[h]
\includegraphics[width=1\linewidth]{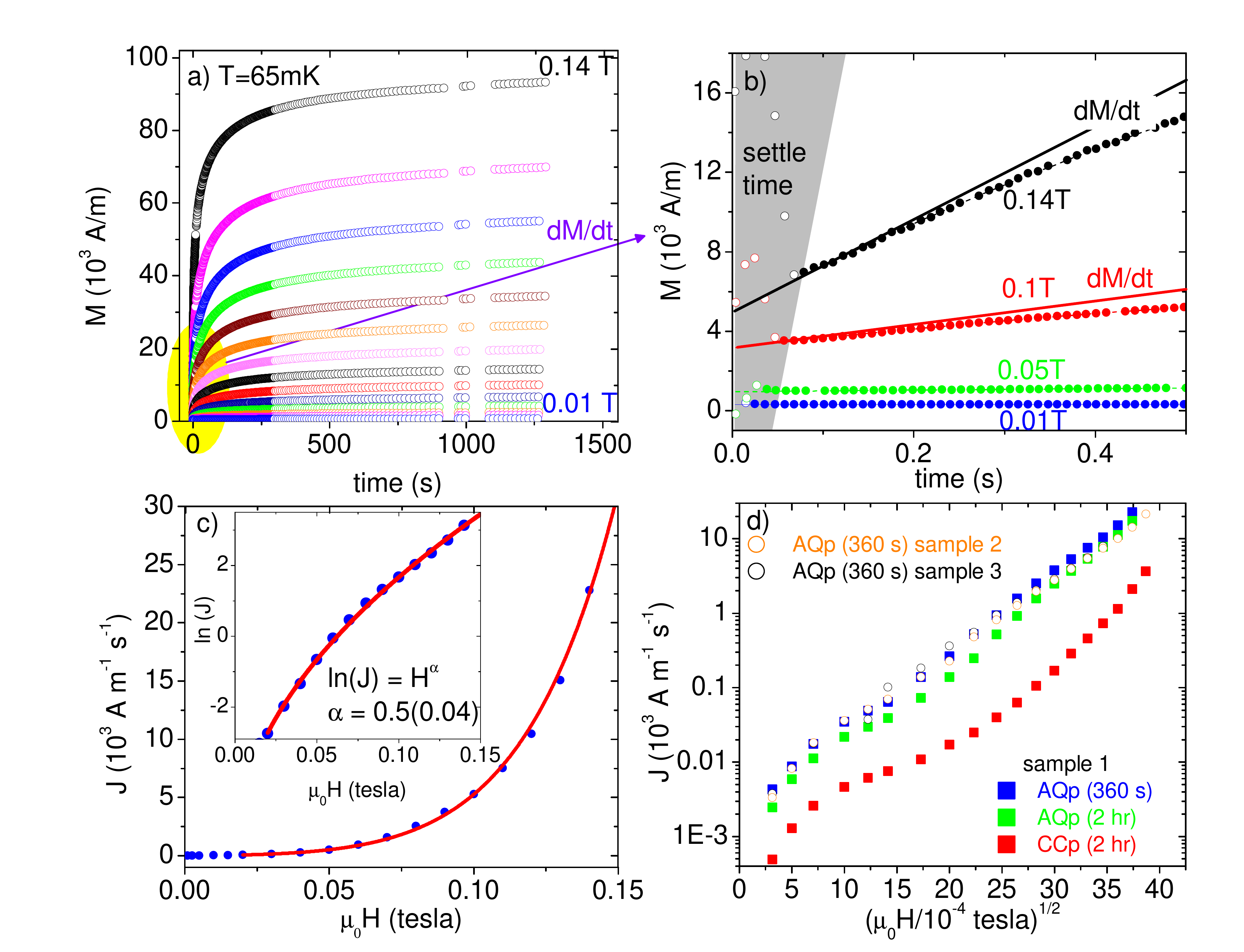}
\caption{Magnetricity in spin ice \DyT~at 65 mK arising from nonequilibrium populations of magnetic monopoles. Fig. 1a: Magnetisation versus time as a function of applied field  after the sample was first prepared using the AQp with a wait period 360 seconds before the field was applied. 
The field values are from bottom to top $\mu_0 H$= 0.001, 0.0025, 0.005, 0.0075, 0.01, 0.015 T and then 0.02 to 0.14 T in steps of 0.01 T. Note $M$ vs time for the low field values fall on top of each other on the scale used in this plot.
Fig. 1b:  Magnification of four relaxation curves.  The dashed lines are polynomial fits to the data points below 1.5 s  but without using the data taken during the field ramp, the open circles.  Two examples of the derivative of the fit  $\partial M/\partial t$ are shown by the straight full lines, and were evaluated at the point in time immediately after the field has stabilised. Fig. 1c: Magnetic current density $J$ as a function of applied field $H$ at 65~mK extracted from the data of Fig. 1a. Inset: $\log J$ versus $H$ for $\mu_0 H\ge0.02$ T.  The line is a free fit to the data points to determine the exponent $\alpha$ of $H$ in $J \sim \exp(H^\alpha)$.  The observed $\alpha \approx 1/2$ is characteristic of the unbinding of pairs of magnetic monopoles that interact according to Coulomb's law of force. 
Fig. 1d: The effect of cooling,  waiting time and sample variation on $J$. For sample 1 the CCp data were cooled at the slowest rate, resulting in a much smaller initial density of monopoles and thus smaller monopole current. The difference between the two AQp curves  taken with two different wait times for sample 1 indicate that even at 65 mK, monopoles gradually recombine as a function of time. The figure also displays data for two other samples using the AQp with a wait period 360 seconds before the field was applied. Note sample (3) had the $\sim 10$\% nuclear spins removed. All samples behave qualitatively similar. }
\label{one} 
\end{figure}

\begin{figure}[h]
\includegraphics[width=1\linewidth]{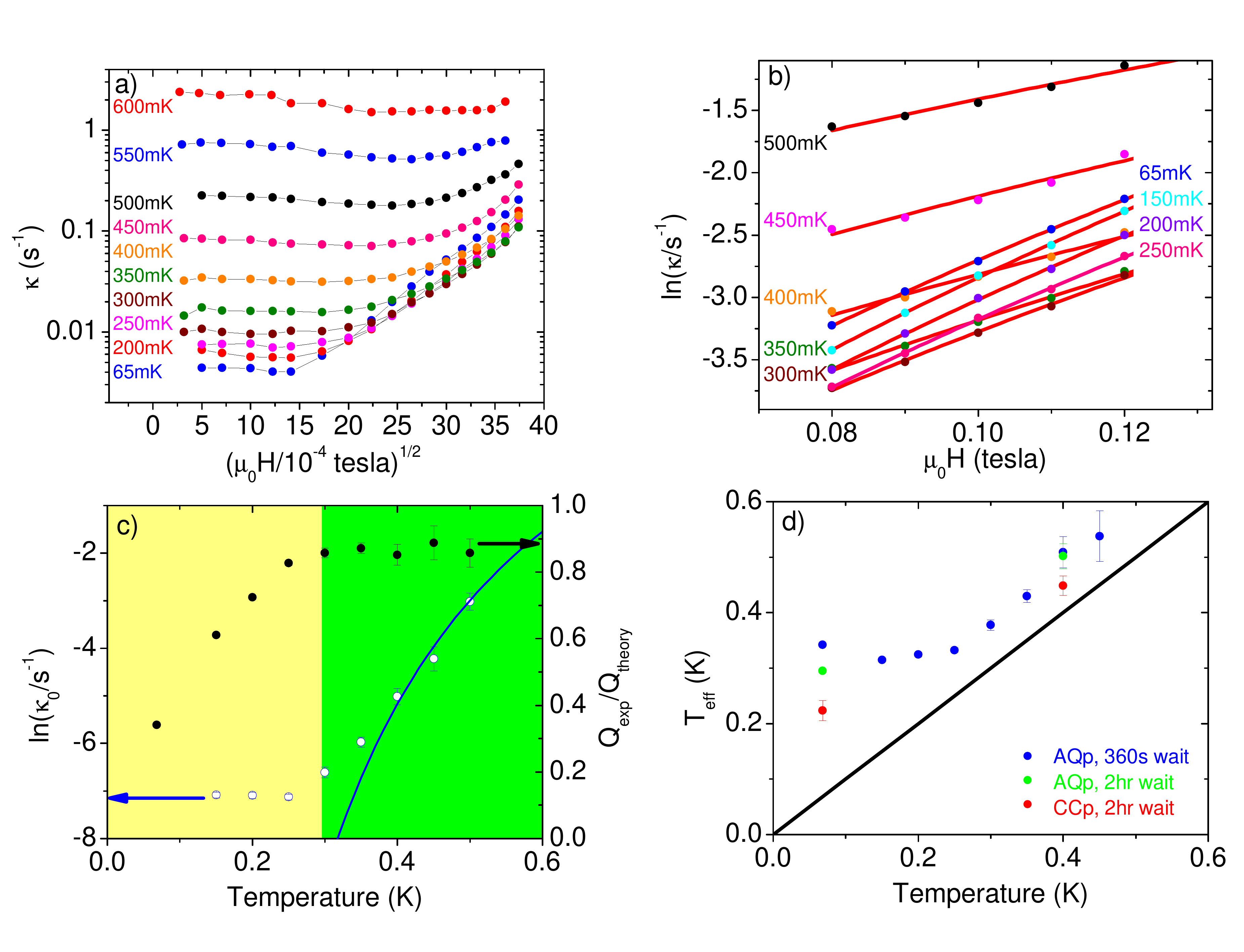}
\caption{Ohmic and non-Ohmic monopole conductivity $\kappa$ as a function of magnetic field. Fig. 2a: Field and temperature dependence  for the AQp prepared sample with a 360 s wait period.  The roughly constant $\kappa$ at low fields implies Ohmic conduction, while at higher fields the conduction is non--Ohmic. Fig. 2b: An approximate analysis of the high field data ($\mu_0 H=$ 0.08 T - 0.12 T, points) using Onsager's {\it unscreened} Wien effect expression,  $\kappa = \kappa_0 \sqrt{F(H)}$ (the conservative field range used for the fits is discussed in the Methods and SI1).
Fig. 2c:  The fitted parameters $Q_{\rm exp}(T)$ and $\kappa_0(T)$ agree with theory at $T \ge 0.3$ K only (green shaded region where $Q_{\rm exp}/Q_{\rm theory} \approx 1$, $\kappa_0 \approx 300 \exp(-4.35/T)$ s$^{-1}$, blue line), with increasing departures from theory at $T <0.3$ K (yellow shaded region). This suggests a restoration of quasiparticle kinetics at $T \ge 0.3$ K.  Fig. 2d: Alternative parameterisation with the temperature $T$ replacing $Q$ as the fitted parameter. Blue points as in Fig. 2a, b, green points have a 2 h wait time and red points are CCp; the black line is the set temperature. The error bars are calculated from the experimental standard deviation.
}   
\label{two}
\end{figure}

\begin{figure}[h]
\includegraphics[width=0.99\linewidth]{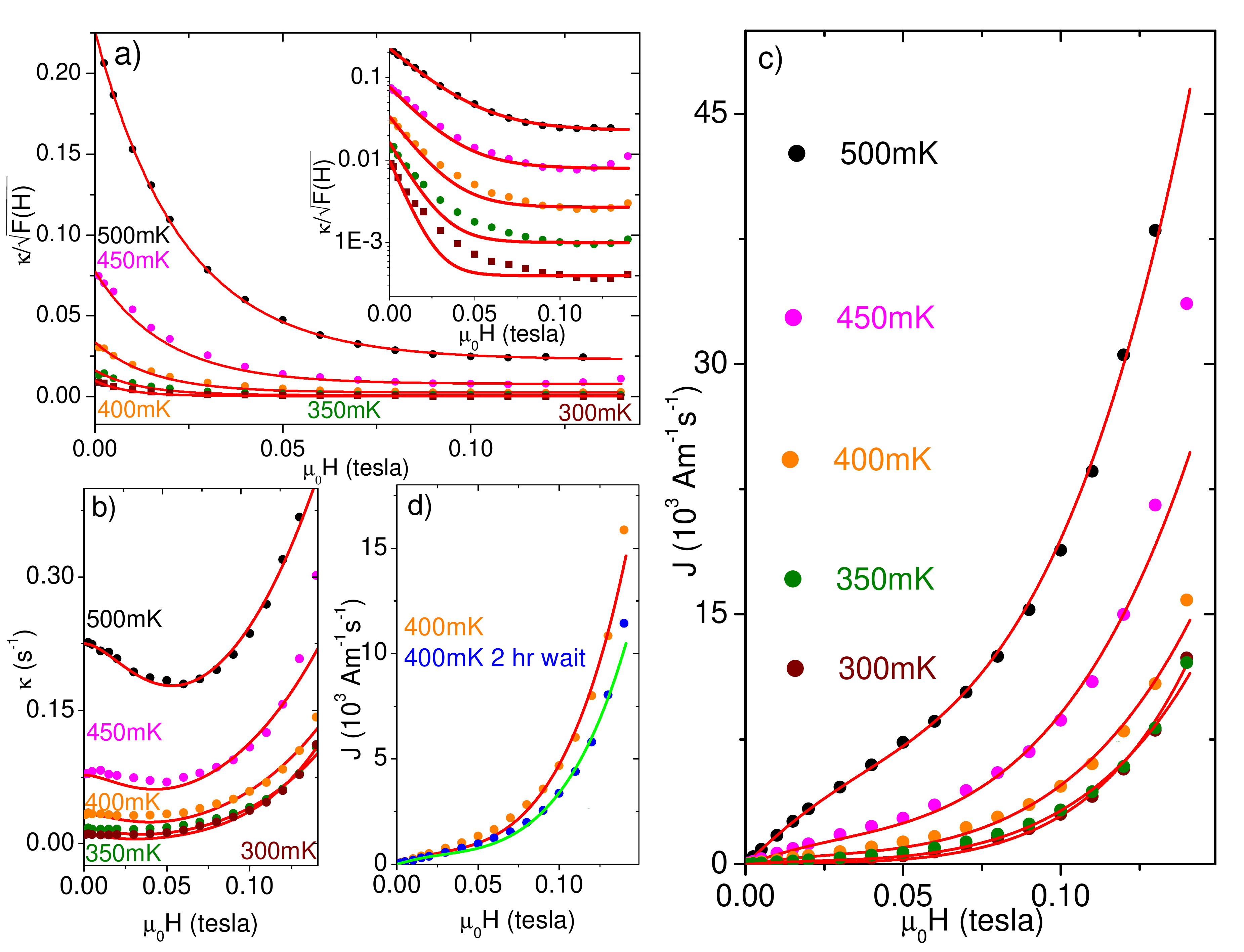}
\caption{High field Wien effect for magnetic monopoles in spin ice, fitted using the theoretical charge~\cite{CMS} and  accounting for charge screening as in Refs.~\citenum{Nmat,Kaiser}: data at $> 0.12$ T are displayed but not fitted (see Methods and SI1). Fig. 3a:  Reduced monopole conductivity $\tilde{\kappa}$ at 300-500 mK, versus field, where $\tilde{\kappa}$ is the monopole conductivity divided by that predicted by Onsager's unscreened theory (Methods). The exponential approach to a finite asymptote confirms the high field Wien effect in spin ice~\cite{Nmat,Kaiser}. The exponential decay closely follows the screening theory of Ref.~\citenum{Kaiser} (full line), albeit with an anomalously small fitted activity coefficient, $\gamma_0$. Inset: same, on a logarithmic scale. Fig. 3b: corresponding conductivity, $\kappa$ versus field. Fig. 3c: the current density, $J$ versus field -- the nonlinearity clearly illustrates deviations from Ohm's law that are accounted for by a strongly screened Wien effect.  Fig. 3d: equivalent data for 400 mK for different wait times, as indicated on the figure. The functional form of $J(H)$ is not severely affected by wait time.
}   
\label{three}
\end{figure}

\begin{figure}[h]
\includegraphics[width=.85\linewidth]{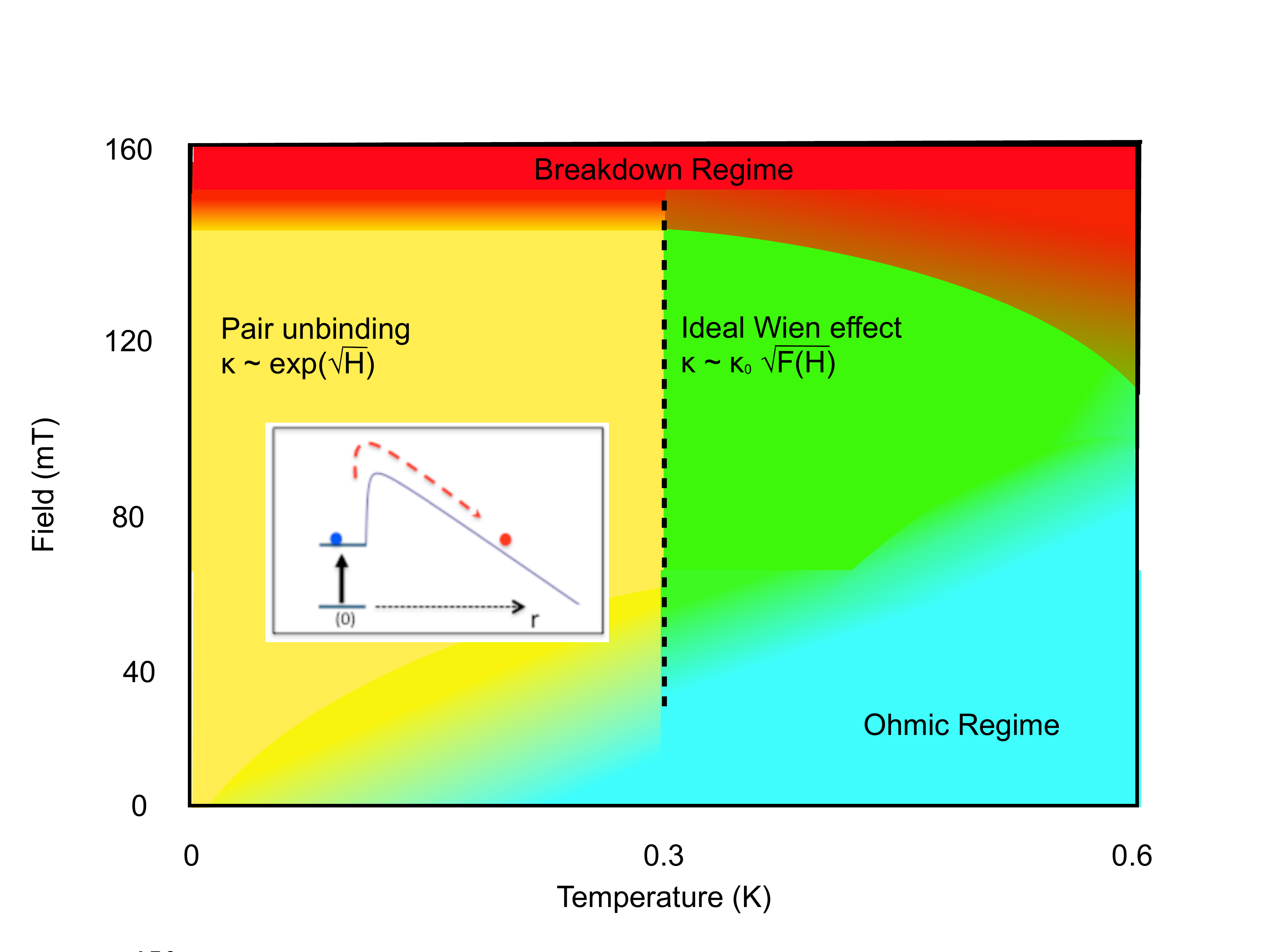}
\caption{Far-from equilibrium magnetic monopole conductivity in spin ice: schematic of our main results for the short-time monopole conductivity as a function of temperature and magnetic field. There are four regimes: (i) a regime of Ohmic conduction at low field (blue), (ii) a regime of the ideal Wien effect for magnetic monopoles (green, $>0.3$ K), (iii) a regime of metastable pair unbinding, where the observed characteristic $\exp(\sqrt{H})$ conductivity is a direct experimental verification of the pairwise Coulomb interaction between magnetic monopoles (yellow, $< 0.3$ K), and (iv) a regime of breakdown or thermal runaway (red, see SI1). The inset shows the field-enhanced unbinding of a metastable (thermally-quenched) monopole-antimonopole pair over the Coulomb barrier as a function of distance ($r$).  Thermal generation of such pairs from the monopole vacuum (0), (vertical arrow) sets in above $\sim 0.3$ K. 
}   
\label{four}
\end{figure}

\end{document}